\documentclass[12pt]{article}

\usepackage[margin=1.5cm]{geometry}

\usepackage{amsmath,amsfonts,mathtools}
\usepackage{microtype}
\usepackage{hyperref}
\usepackage{color}
\usepackage{bm}

\usepackage{tikz}

\begin{document}

	
	
	\title{Stochastic niche-based models for the evolution of species~\thanks{We would like to thank the Australian Research Council for funding this research through Discovery Project DP180100352.}}
	


	\author{
		Albert Ch. Soewongsono\thanks{Discipline of Mathematics, University of Tasmania, Australia, email: albert.soewongsono@utas.edu.au}
		\ and
		Barbara R. Holland\thanks{Australian Research Council Centre of Excellence for Plant Success, Discipline of Mathematics, University of Tasmania, Australia, email: barbara.holland@utas.edu.au} 
		\ and
		Ma{\l}gorzata M. O'Reilly~\thanks{Discipline of Mathematics, University of Tasmania, Australia, email: malgorzata.oreilly@utas.edu.au}
	}

	\date{\today}
	\maketitle
	
	\section{Introduction}
	
	There have been many studies to examine whether one trait is correlated with another trait across a group of present-day species. For example, a question of interest could be: Do species with larger brains tend to have longer gestation times? In 1985, Felsenstein~\cite{felsenstein1985phylogenies} published a key paper arguing that any such comparisons need to account for correlations that arise due to shared evolutionary ancestry (as encapsulated by a phylogenetic tree). This paper prompted great interest in the development of phylogenetic comparative methods to model statistical dependence among species.
	
	Since the introduction of the phylogenetic comparative method some authors have questioned its validity and utility. For example, Westoby et al. \cite{westoby1995misinterpreting} argue that it is not a simple task to disentangle the effects of ecological viability and phylogenetic history. In order to address these issues, it is necessary to have a biologically realistic model to generate evolutionary trees, that incorporates information about the ecological niche occupied by species.
	
	Price presented a simple model along these lines in~\cite{price1997correlated}. Price defined a two-dimensional niche space formed by two continuous-valued traits, in which new niches arise with trait values drawn from a bivariate normal distribution. When a new niche arises it is occupied by a descendant species of whichever current species is closest in ecological niche space. In the sequence of speciation events, more species are then evolved from already-existing species to which they are ecologically closest.
	
	Here, we explore ideas that extend Price's adaptive radiation model in~\cite{price1997correlated}. The first extension is to increase the dimensionality of the niche space by considering more than two continuous traits. The second extension is to allow both extinction of species (which may leave unoccupied niches) and removal of niches (which causes species occupying them to go extinct).
	
	To build our model for the evolution of traits and species, we first consider a continuous-time stochastic process with state space $\mathcal{S} = \{\left(N(t),{\bf Z}(t),{\bf y}(t)\right):t \geq 0\}$ where
	\begin{itemize}
		\item $N(t)=0,1,2,\ldots$ is the total number of niches that were born by time $t$; 
		\item the matrix ${\bf Z}(t)=[{\bf z}_i(t)]_{i=0,\ldots,N(t)}$ is given by the vectors ${\bf z}_i(t)=[z_i^{(1)}(t),\ldots,z_i^{(m)}(t)]$ recording the values of the $m$ environmental parameters of niche $i$ in the niche space with $m$ traits; and 
		\item the vector ${\bf y}(t)=[y_i(t)]_{i=1,\ldots,N(t)}$, is such that $y_i(t)\in\{-1,0,\ldots,N(t)\}$, where $y_i(t)=-1$ indicates that niche $i$ is no longer available to be occupied, $y_i(t)=0$ indicates that niche $i$ is not currently occupied, and $y_i(t)>0$ indicates niche $i$ is currently occupied; where $y_i(t)\in\{1,\ldots,N(t)\}\setminus \{i\}$ is the parent niche from which niche $i$ was occupied. 
	\end{itemize}  
	
	This process implicitly defines a phylogeny. To explore if trees generated under such a model (or under different parametrizations of the model) are realistic we can compute a variety of summary statistics that can be compared to those of empirically observed phylogenies. For example there are existing statistics that aim to measure: tree balance \cite{aldous2001stochastic}, the relative rate of diversification \cite{pybus2002}, and phylogenetic signal of traits \cite{freckleton2015phylogenetic,pagel1999inferring}.

	
	\section{The model: State space}
	
	We build on the above initial idea as follows. For a more realistic model, let 
	$$\mathcal{B}=\{(z^{(1)},\ldots,z^{(m)}):z^{(i,min)}\leq z^{(i)}\leq z^{(i,max)}\} \subset \mathbb{R}^m$$
	corresponds to the fundamental niche with $m$ parameters that the species may tolerate, such as the minimum and maximum temperature, or nitrogen content of the soil, or precipitation. Note that in this paper, we assume that all species share the same fundamental niche $\mathcal{B}$.

	When a new niche appears in some geographical location, then the environmental parameters at each point in this new location, will be within the fundamental niche $\mathcal{B}$.
	
	Therefore, we assume that niche $i=1,\ldots,N(t)$ at time~$t$, occupies some geographical location denoted $\mathcal{A}_i(t) \subset \mathbb{R}^2$, such that at each point $(x_1,x_2)\in \mathcal{A}_i(t)$, all environmental parameters ${\bf z}_{i;x_1,x_2}(t)=[z_{i;x_1,x_2}^{(1)}(t),\ldots,z_{i;x_1,x_2}^{(m)}(t)]$ are within the fundamental niche, which we write as
	$${\bf z}_{i;x_1,x_2}(t)=[z_{i;x_1,x_2}^{(1)}(t),\ldots,z_{i;x_1,x_2}^{(m)}(t)]
	\in\mathcal{B}.$$
	We denote
	$$
	\mathcal{A}(t)=(\mathcal{A}_i(t))_{i=1,\ldots,N(t)}
	$$
	and define 
	$$\mathcal{Z}(t)=({\bf z}_{i;x_1,x_2}(t))_{(x_1,x_2)\in\mathcal{A}_i(t);i=0,\ldots,N(t)}$$
	which records the geographical locations of the $N(t)$ niches and the values of the $m$ environmental parameters for niche $i$ at location $(x_1,x_2)\in\mathcal{A}_i(t)$, at time $t$.

	That is, we consider a continuous-time stochastic process $\{\left(N(t),\mathcal{A}(t),\mathcal{Z}(t),{\bf y}(t)\right):t \geq 0\}$ with the variable 
	$$N(t)\in\{0,1,2\ldots\}$$ 
	recording the number of niches available at time $t$; the variable 
	$$\mathcal{A}(t)\subset (\mathbb{R}^2)^{N(t)}$$ recording the geographical locations of each of the $N(t)$ niches; the variable $$\mathcal{Z}(t)\subset \mathcal{B}^{N(t)}$$ 
	recording the values of the $m$ environmental parameters of each niche at each possible location $(x_1,x_2)$ at time $t$; and the variable $${\bf y}(t)\in \{-1,0,\ldots,N(t)\}^{N(t)}$$ 
	recording the information about whether a niche is occupied or not, and if it is occupied, what is its parent niche.
	
	Therefore, the state space of such defined process is,
	\begin{eqnarray*}
		\mathcal{S}&=&
		\Big\{\left(N,\mathcal{A},\mathcal{Z},{\bf y}\right):N\in\{0,1,2\ldots\},
		\mathcal{A}\subset (\mathbb{R}^2)^{N},
		\\
		&&
		\quad
		\mathcal{Z}\subset {\mathcal{B}}^{N},
		{\bf y}\in \{-1,0,\ldots,N\}^{N}\Big\}.
	\end{eqnarray*}

	\section{Niche process}

	In order to study the niche process on some map of geographical locations within $\mathbb{R}^2$, we assume that the map is a grid divided into $J\times K$ rectangular patches $(x_1,x_2)$, $x_1=1,\ldots,J$, $x_2=1,\ldots,K$. 
	
	We will say that a patch $(x_1,x_2)$ is suitable when its parameters are inside the fundamental niche, that is, 
	$${\bf z}(x_1,x_2)=[z^{(1)}(x_1,x_2),\ldots,z^{(m)}(x_1,x_2)]\in\mathcal{B}.$$
	In this paper, we will refer to these combinations of suitable environment and a particular location (patch) as a niche.

	Each suitable patch (a niche) may be occupied or not. We note that $J\times K$ is the maximum possible number of niches on the map. We illustrate this in Figure~\ref{fig:Grid}.

	\begin{figure}[h!]
		\begin{center}
			\resizebox{0.4\textwidth}{!}{
				\begin{tikzpicture}
					\foreach \i in {1,...,4} {
						\draw [very thin,gray] (\i,1) -- (\i,6)  node [below] at (\i+0.5,1) {$\i$};
						\draw [very thin,gray] (5,1) -- (5,6);
					}
					\foreach \i in {1,...,5} {
						\draw [very thin,gray] (1,\i) -- (5,\i) node [left] at (1,\i+0.5) {$\i$};
						\draw [very thin,gray] (1,6) -- (5,6);
					}			
					\draw[fill=black!30] (1,1) rectangle (2,2);
					\draw[fill=black!30] (4,4) rectangle (5,5);
					\draw[fill=black!30] (3,2) rectangle (4,3);
					\draw[fill=black!30] (2,5) rectangle (3,6);
					\draw[fill=black!30] (1,3) rectangle (2,4);
					\node[black] at (2.5,1.5){0};
					\node[black] at (3.5,1.5){1};
					\node[black] at (4.5,1.5){1};
					\node[black] at (2.5,2.5){0};
					\node[black] at (1.5,2.5){0};
					\node[black] at (4.5,2.5){1};
					\node[black] at (4.5,3.5){0};
					\node[black] at (3.5,3.5){0};
					\node[black] at (2.5,3.5){1};
					\node[black] at (2.5,4.5){1};
					\node[black] at (1.5,4.5){1};
					\node[black] at (3.5,4.5){1};
					\node[black] at (1.5,5.5){0};
					\node[black] at (3.5,5.5){1};
					\node[black] at (4.5,5.5){0};
				\end{tikzpicture}
			}
		\end{center}
		\caption{An example of a two-dimensional map indicating the location of each patch, with $J \times K=4 \times 5=20$ patches. Each shaded rectangle indicates that the patch is not suitable to be inhabited, while white rectangles indicate that the patch is suitable. $1$ and $0$ denote whether a patch is occupied or unoccupied, respectively.		
		}
		\label{fig:Grid}
	\end{figure}

	\subsection{Niche process: Model I}
	
	Under Model I, we describe the niche process as follows.

	Suppose that new niches appear on the grid at a rate $\lambda(t)$ of some time-dependent Poisson Process. Given a new niche appears on the grid (an arrival occurs in the Poisson process with the rate $\lambda(t)$), it will be at a location $(x_1,x_2)$ according to some probability $\mathbb{P}(x_1,x_2)$.
	
	As example, we may assume that a new niche appears in a completely new location separate from other existing niches with probability~$p$, or in a location neighboring to one of the existing niches with probability $(1-p)$.
	
	Additionally, we may assume that the choice of the location of the new niche is then performed in a uniform manner. That is, if there are $N$ potential niches in separate locations, then the probability of choosing a niche at location $(x_1,x_2)$ is $\mathbb{P}(x_1,x_2)=p/N$. Similarly, if there are $M$ potential niches in neighboring locations, then the probability of choosing a niche at a neighboring location $(x_1,x_2)$ is $\mathbb{P}(x_1,x_2)=(1-p)/M$.
	
	Furthermore, once the location $(x_1,x_2)$ is chosen, the values ${\bf z}(x_1,x_2)$ from the fundamental niche $\mathcal{B}$ are drawn from some distribution, $F({\bf z}(x_1,x_2))
	$, defined for all ${\bf z}(x_1,x_2)\in\mathcal{B}$. This distribution could be based on empirical data.
	
	Finally, we assume that niches disappear from the grid at a rate $\mu(t)$ of another time-dependent Poisson Process. Given a niche disappears from the grid (an arrival occurs in the Poisson process with the rate $\mu(t)$), it will be at a location $(x_1,x_2)$ according to some probability $\widetilde {\mathbb{P}}(x_1,x_2)$.
	
	For example, the choice of niche could be performed in a uniform manner. Once the niche at $(x_1,x_2)$ is chosen, the values ${\bf z}(x_1,x_2)$ at that niche are drawn from some distribution $\widetilde F({\bf z}(x_1,x_2))$, defined for all ${\bf z}(x_1,x_2)\in (\mathbb{R}^m\setminus \mathcal{B})$.

	\subsection{Niche process: Model II}

	Under Model II, we describe the evolution of ${\bf z}(x_1,x_2)$ which then drives the niche process, as follows.
	
	We assume that the process $\{z^{(k)}(t):t\geq 0\}$ which describes the evolution of the parameter $k=1,\ldots,m$, at any patch $(x_1,x_2)$, is modeled using the Ornstein--Uhlenbeck (OU) process  described in \cite{butler2004phylogenetic}, that is,
	$$
	dz^{(k)}(t)=\alpha^{(k)}\lceil \theta^{(k)} - z^{(k)}(t)\rceil dt + \sigma^{(k)} dB^{(k)}(t),
	$$
	where $dB^{(k)}(t)$ are independent and identically distributed (iid) random variables drawn from normal distribution with mean $0$ and variance $dt$, $\theta^{(k)}$ is the optimum value for parameter $k$, $\alpha^{(k)}$ and $\sigma^{(k)}$ are the ``pull" and dispersion coefficients for parameter $k$, respectively.
	
	Then, if the values $z^{(k)}(x_1,x_2)$ of all parameters $k=1,\ldots,m$ at patch $(x_1,x_2)$ at time $t$ are such that ${\bf z}(x_1,x_2)\in\mathcal{B}$, then patch $(x_1,x_2)$ is considered to be a niche (a suitable patch) at time $t$.
	
	Alternatively, in the case the parameter values ${\bf z}(x_1,x_2)$ leave the fundamental niche $\mathcal{B}$ at time $t$, so that ${\bf z}(x_1,x_2)\notin\mathcal{B}$, then niche $(x_1,x_2)$ is removed and becomes a patch $(x_1,x_2)$ that is no longer suitable at time $t$.

	\subsection{Niche process: Model III}
	
	Under Model II, we also assume that the evolution of the parameters ${\bf z}(x_1,x_2)$ drive the niche process. This time however, we model the evolution of ${\bf z}(x_1,x_2)$ using a different process, so as to take into account potential interactions between the parameter values at nearby locations.
	
	For notational convenience, let the column vector
	$$
	{\bf z}_{\ell}(t) =\left[z_{\ell}^{(1)}(t);\cdots;z_{\ell}^{(m)}(t)\right],
	$$ 
	record the parameters values for a patch at some location $\ell = (x_1,x_2)$ at time $t$, for all $\ell=1,\ldots, J \times K$. Patch $\ell$ is a niche (a suitable patch) at time $t$ if and only if ${\bf z}_{\ell}(t)\in\mathcal{B}$.
	
	We note that the parameter $k$ values at any location may not fall outside some interval $[a^{(k)},b^{(k)}]$, where $a^{(k)}$ and $a^{(k)}$ are the lower and upper boundary of possible values, respectively, due to the realistic conditions. That is, 
	$$\mathcal{B}\subset
	[a^{(1)},b^{(1)}]\times\cdots\times[a^{(m)},b^{(m)}].$$
	Let $g_{a,b}(x)$ be a function taking values $g_{a,b}(x)=x$ for $x\in[a,b]$, $g_{a,b}(x)=a$ for $x< a$, and $g_{a,b}(x)=b$ for $x> b$.

	We assume that there is an underlying continuous-time Markov chain (CTMC) $\{W(t):t\geq 0\}$ with a finite state space $\mathcal{D}$ and generator ${\bf T}$, which drives the evolution of the parameters $k=1,\ldots,m$ as follows.
	
	The value of parameter $k$ at location $\ell$ evolves according to the second order fluid model with reflecting boundaries discussed in \cite{gribaudo2008second}, that is, for all $k=1,\ldots,m$, and $t'>t$,
	\begin{eqnarray*}
		z_{\ell}^{(k)}(t')
		&=&
		g_{a^{(k)},b^{(k)}}\Big(z_{\ell}^{(k)}(t) 
		\\
		&&+ N\left((t'-t)\mu^{(k)}_{W(t),\ell}\ ,\ (t'-t){\Sigma^2}^{(k)}_{W(t),\ell}\Big)
		\right)
	\end{eqnarray*}
	where $\mu^{(k)}_{j,\ell}$ is the drift and ${\Sigma^2}^{(k)}_{j,\ell}$ is the variance of parameter~$k$ at location $\ell$ when $W(t)=j$, defined for all $j\in\mathcal{D}$, as
	\begin{eqnarray*}
		\mu^{(k)}_{j,\ell} &=& \sum_{\ell'}\varphi_{\ell'\ell}r^{(k)}_{j,\ell'}
		=r^{(k)}_{j,\ell}+
		\sum_{\ell'\not=\ell}\varphi_{\ell'\ell}r^{(k)}_{j,\ell'},\\
		{\Sigma^2}^{(k)}_{j,\ell} &=& \sum_{\ell'}\varphi_{\ell'\ell}V^{(k)}_{j,\ell'}=V^{(k)}_{j,\ell}
		+
		\sum_{\ell'\not=\ell}\varphi_{\ell'\ell}V^{(k)}_{j,\ell'}
		,
	\end{eqnarray*}
	where $r^{(k)}_{j,\ell}$ and $V^{(k)}_{j,\ell}$ is the drift and the variance of parameter~$k$ at location $\ell$ if it did not depend on other locations; and where $\varphi_{\ell\ell}=1$, and for $\ell'\not=\ell$, $\varphi_{\ell'\ell}\in(0,1)$ are some weights that model the influence of location $\ell'$ on the evolution of parameters at location $\ell$. The function $g_{a^{(k)},b^{(k)}}(\cdot)$ ensures that the parameter values stay within the possible thresholds.

	Under this model, when $W(t)=j$, then the value of parameter $k$ at location $\ell$ is driven by a Brownian motion with drift $\mu^{(k)}_{j,\ell}$ and variance ${\Sigma^{2}}^{(k)}_{j,\ell}$. Both drift and variance change according to the current state in the underlying CTMC $\{W(t):t\geq 0\}$, which drives the evolution.
	
	\section{Species extinction process}
	
	We assume that the extinction of the species at location $\ell$ may occur due to the following two scenarios.
	
	If a niche disappears from the location~$\ell$ on the grid for the duration of at least $\Delta t>0$, then the species present in that location becomes extinct.
	
	Alternatively, the species present at location $\ell$ may become extinct due to some other underlying process that drives the evolution (see the age-dependent extinction model introduced in \cite{alexander2016quantifying}).
	
	\section{Speciation process}

	If a new niche appears at location $\ell$ on the grid (according to one of Models I-III), then speciation may occur when some species moves from some location $\ell'\not= \ell$ to location $\ell$. We assume that this occurs under the following assumptions.
	
	We build on some ideas for a discrete-time metapopulation model studied by McVinish et al. in~\cite{mcvinish2016metapopulation}, to define rates $\gamma_{\ell'\ell}(t)$, in the continuous-time model considered here, at which at time $t$ species inhabit an unoccupied niche at location $\ell$ from location $\ell'\not= \ell$, for all $\ell',\ell=1,\ldots,J\times K$.

	Let $X_{\ell}(t)=0$ when patch $\ell$ is unoccupied and suitable, that is with ${\bf z}_{\ell}(t)\in\mathcal{B}$. We let $X_{\ell}(t)=1$ otherwise. Let $s_{\ell'\ell}$ be the similarity metric of patch $\ell'$ and $\ell$, for all $\ell',\ell=1,\ldots,J\times K$. As an example, if we let $s_{\ell'\ell}=1/d_{\ell'\ell}$ where $d_{\ell'\ell}$ is the distance between the patches $\ell'$ and $\ell$, then the patches that are in closer locations, are considered to be more similar.

	We define the rate at which niche $\ell$ becomes occupied from niche $\ell'$ at time $t$ as	
	$$\gamma_{\ell'\ell}(t)=\gamma\times 
	P_{\ell'\ell}(t)
	\times \left(1-X_{\ell}(t)\right),$$
	where $\gamma$ is the  colonisation rate per patch, and
	\begin{eqnarray*}
		P_{\ell'\ell}(t)&=&
		\left\{
		\begin{array}{ll}
			\frac{s_{\ell'\ell}  X_{\ell'}(t)}{\sum_{\ell'} s_{\ell'\ell} X_{\ell'}(t)}&\mbox{if }
			\sum_{\ell'} s_{\ell'\ell} X_{\ell'}(t)\not= 0,\\[2ex]
			0&\mbox{otherwise;}
		\end{array}
		\right.	
	\end{eqnarray*}
	is the probability that niche $\ell$ becomes occupied from niche $\ell'$ given a colonisation event has occurred.

	We note that when $X_{\ell'}(t)=1$ for at least one $\ell'\not=\ell$ and $X_{\ell}(t)=0$, then
	$\sum_{\ell'}P_{\ell'\ell}(t)=1$
	and so
	\begin{eqnarray*}
		\gamma_{\ell}(t)&=&
		\sum_{\ell'\not=\ell}
		\gamma_{\ell'\ell}(t)
		=
		\gamma\times
		\left(1-X_{\ell}(t)\right)
		=\gamma.
	\end{eqnarray*}

	Therefore, the colonisation rate $\gamma$ is the parameter that controls the speciation, which is interpreted as the rate at which an unoccupied niche $\ell$ becomes occupied.

	On the other hand, when either $X_{\ell}(t)=1$, or $X_{\ell'}(t)=0$ for all $\ell'\not=\ell$ and $X_{\ell}(t)=0$, then $\gamma_{\ell}(t)=
	0$.
	
	In our future paper, we will apply the above modelling ideas to study correlation between traits across present-day species by taking into account their phylogenetic tree generated using the above models.

	\bibliographystyle{abbrv}
	\bibliography{publish3_biblio}
\end{document}